\documentclass[aps,pra,reprint,superscriptaddress,twocolumn,groupedaddress,showpacs,amsmath,amssymb,amsfonts]{revtex4}
\usepackage{graphicx}
\usepackage{bm}
\usepackage{color}

\def\uband{13.5$\pm$1\%\;nm\;}
\def\eband{11.2$\pm$1\%\;nm\;}
\def\kn{\underline{k\vphantom{k}}}\def\kv{\overline{k}\vphantom{k}}
\def\nn{\underline{n\vphantom{n}}}\def\nv{\overline{n}\vphantom{n}}
\def\gn{\underline{g}\vphantom{g}}\def\gv{\overline{g}\vphantom{g}}

\begin {document}
\title{Extreme-ultraviolet light source for lithography based on an expanding jet \\ of dense xenon  plasma supported by microwaves}
\author{I. S. Abramov}
\author{E. D. Gospodchikov}
\author{A. G. Shalashov}
\email{ags@appl.sci-nnov.ru}
\affiliation{Institute of Applied Physics of the Russian Academy of Sciences, 46 Ulyanova str., 603950 Nizhny Novgorod, Russia}


\begin{abstract}
We discuss a concept of a pointlike source of extreme-ultraviolet (EUV) light based on a nonequilibrium microwave discharge in an expanding jet of dense xenon plasma with multiply charged ions. The conversion efficiency of microwave radiation to EUV light is calculated, and physical constraints and opportunities for future devices are considered. Special attention is given to trapping of spontaneous line emission inside a radiating plasma spot that significantly influences the efficiency of an EUV light source. 
\end{abstract}


\maketitle

\section{\label{sec:intro}Introduction}
Transition to exposure using  extreme-ultraviolet (EUV) radiation is vital for the development of next-generation projection lithography \cite{bakshi_2006}. The only realistic method of EUV light generation is based on line radiation of multiply charged ions considering that stripping causes a shift of the ion emission spectrum toward shorter wavelengths for highly ionized charge states. In the EUV band, radiation is efficiently absorbed while propagating in the atmosphere and by elements of refractive optics \cite{wagner_2010}. Thus, EUV radiation should be manipulated in vacuum conditions and with reflective optics, such as multilayer mirrors. These mirrors have a narrow wavelength band where their reflection is high enough for efficient focusing of EUV radiation. Particular mirror materials and technology define  the exact band of EUV radiation sources required by industry, and, consequently, the kind of emitting elements.

Most of the existing EUV sources operate at \uband\!,  corresponding to peak reflection coefficients of Mo/Si mirrors \cite{andreev_2003}. For this band, tin ions are an adequate choice due to a significant number of strong radiation lines at \uband for Sn$^{7+}$--Sn$^{12+}$. The most-successful projects by ASML Cymer and Gigaphoton, which are  almost ready for industry solutions, use evaporation of tin droplets in the focused beam of a CO$_2$ laser. 
The Gigaphoton source has an average EUV power above 100 W during 22 h of stable operation at 50 kHz  and a conversion efficiency up to 5\% \cite{giga_mizoguchi_2017}. The ASML machine \cite{asml_schafgans_2015} currently operates at 205 W, and developers are confident of achieving 250 W in  2018 \cite{moore}. That wattage  saturates the current demand of  microelectronics; however, the mid-term development of lithography (a transition to less than 5-nm node)  requires an EUV source of up to 1000\,W,  which, experts believe, is nearly impossible to realize with a laser-produced plasma \cite{moore}.

Another approach involving use of a microwave discharge, and that is potentially less constrained in terms of EUV power, is under development at the Intitute of Applied Physics of the Russian Academy of Sciences.   
Microwave discharge has a number of other advantages over laser-produced plasmas, such as simpler overall design and safe operation for EUV collecting optics, which are saved from being spoiled by solid target particles and fast ions typical of explosive (of few nanoseconds) laser-produced plasmas. The duration of microwave pulses is longer [from tens of microseconds to continuous-wave (cw) operation], and there are no channels for significant ion heating due to direct resonant power load into electrons.  In pioneer experiments, a strongly nonequilibrium tin plasma was confined in an open magnetic trap and heated by high-power microwaves, resulting in up to 50-W emission in the \uband band \cite{Vodopyanov,ChkhVod,Vodopyanov1}.

Use of heavy noble gases, presumably xenon, instead of tin, opens new opportunities for microwave-based sources.  Xenon ions are not as efficient emitters at \uband as tin ions \cite{churilov_2003}, there is a whole family of strong lines in the \eband band for  Xe$^{10+}$ and a few extra lines for neighboring ions \cite{churilov_2002,churilov_2004,saloman_2004}. For this band, there are Ru/Be and Mo/Be mirrors that have a nearly-twofold-higher peak reflection coefficient in \eband in comparison with Mo/Si mirrors in the \uband band \cite{ckhalo_adv_2013}. Thus, a combination of xenon as the emitter and new mirrors operating at \eband may potentially be more profitable for EUV lithography development.

The first successful experiments aimed at the realization of a pointlike highly emissive discharge in a noble gas have been reported \cite{glyavin_apl_2014,vodopyanov_smp_2017,vodopyanov_IRMMW_2017}. 
Such experiments are possible with recent progress in a development of high-power gyrotrons operating at frequencies of 250--670 GHz and providing a power of 100 kW and higher \cite{subTHz1,subTHz2,subTHz3}. Going to higher frequencies allows a microwave discharge to be supported at higher plasma densities (more than $10^{16}\;$cm$^{-3}$ in the reported experiments), and thus increase in emissivity (up to 10 kW in the 110--180-nm band). In this case, the discharge may be ignited in a gas jet launched from a small nozzle at almost atmospheric pressure, and then it freely expands into a pumped-out chamber (7--200\,mTorr at the wall). 

In this paper, inspired by the success of recent experiments, we report a model predicting the conversion efficiency into the EUV band in the expanding gas jet and study the prospects of xenon-based EUV sources in present technological circumstances.
An essential new feature of xenon discharge is the much higher plasma density(at least two orders of magnitude)  than in tin plasma analyzed earlier \cite{shalashov_jetp_2016, abramov_rf_2016,shalashov_os_2016,abramov_pop_2017}. Correspondingly, we consider new EUV emission physics  taking into account the effects of radiation reabsorption in an optically thick plasma. 


\section{\label{sec:concept}EUV source concept}

\begin{figure}[tb]
\includegraphics[width=85mm]{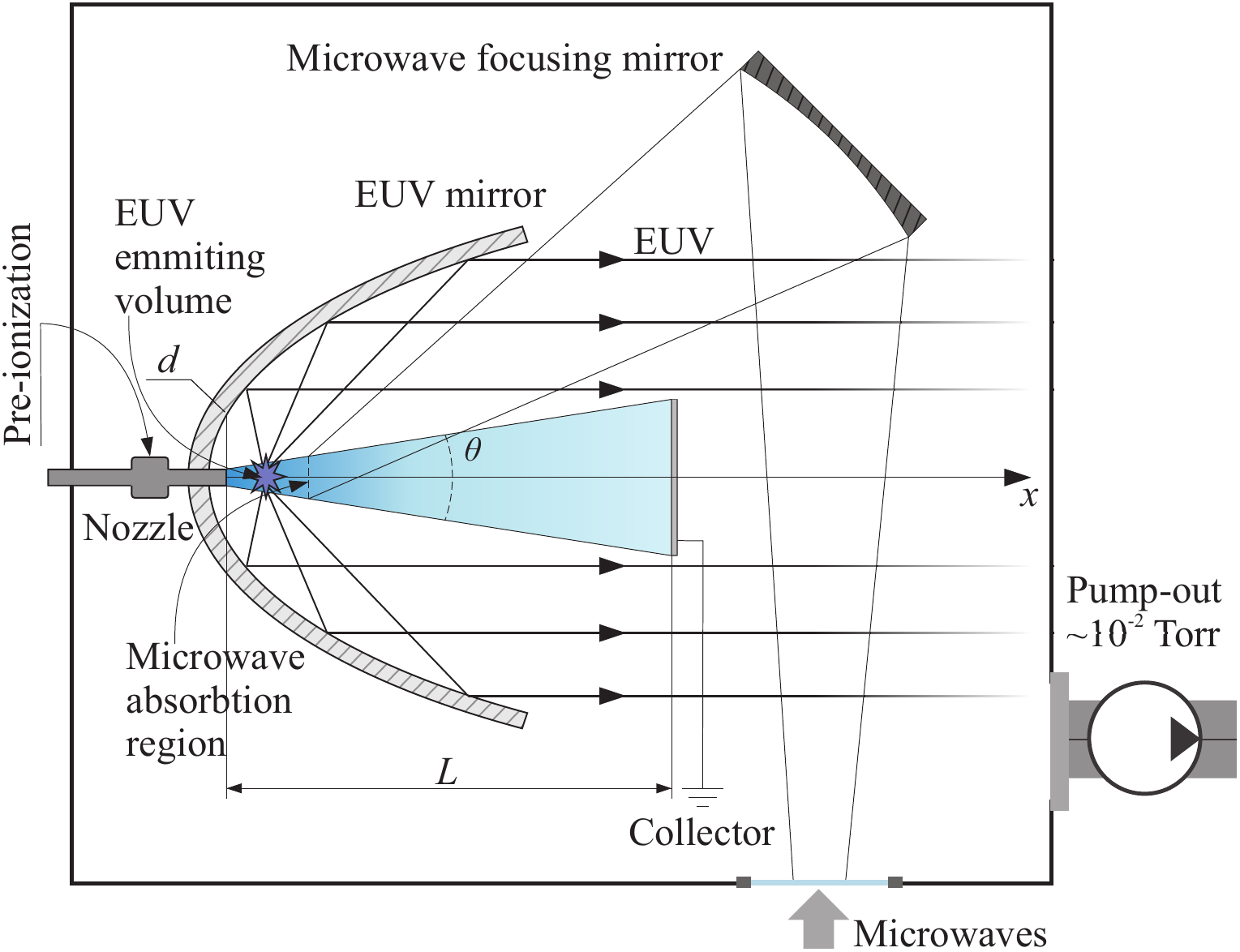}
\caption{Concept of EUV light source as proposed in \cite{glyavin_apl_2014}.}\label{Fig01}
\end{figure}

The concept of an EUV light source with a freely expanding xenon plasma jet is illustrated in Fig.~\ref{Fig01}. 
Xenon is injected into a vacuum chamber with a nozzle going through a paraboloid mirror used for collecting and focusing EUV light. 
High-power microwave radiation from a gyrotron is focused behind the nozzle and is resonantly absorbed by the electron component under conditions of  plasma resonance when the electron Langmuir frequency becomes equal to the wave frequency. This condition is locally met somewhere along the expanding jet as electron density continuously decreases toward the collector. High electron thermal conductivity (typical for the plasma parameters considered) provides efficient heat transport outside the plasma resonance. Thus, the resonant heating may occur  outside the EUV emitting zone (e.g., \ in a less-dense plasma), which allows  the discharge to be supported over a wider range of microwave frequencies.
Direct power load into electrons is beneficial for generation of highly charged ions. Such heating leads to high-electron-temperature nonequilibrium plasmas ($T_e\sim100\;$eV, while the ion temperature $T_i\lesssim1\;$eV)  characterized by high rates of electron-impact ionization and line excitation and suppression of the recombination. 

The setup operates in a quasi-stationary regime, which is more favorable to increase the efficiency of EUV emission. This is possible due to the long duration of microwave pulses for available gyrotrons (50$\,\mu$s or longer)  in comparison with the ion fly time through the chamber (few$\,\mu$s). 
The most critical part of the discharge is related to its start-up, at which slow gas flow initially consisting of neutral species becomes ionized and accelerated up to ion-acoustic velocity \cite{shalashov_jetp_2016, abramov_rf_2016,shalashov_os_2016}. Partial preionization of the flow would essentially decrease the average time that an ion spends in low-charge stages and thus accelerates the transition to the stationary regime within the gyrotron pulse duration. Methods of such preionization are developed and checked experimentally.

Expanding plasma is absorbed by a metal collector located far from the emitting region. The decrease of plasma density with jet expansion provides good localization of the discharge, resulting, in particular, in the formation of a spotlike EUV emitting region near the nozzle (less than 1 mm in all dimensions for the reported experiments \cite{glyavin_apl_2014,vodopyanov_smp_2017,vodopyanov_IRMMW_2017}). The latter allows efficient collection of EUV radiation. 

\section{\label{sec:theo}Theoretical model}

Our model is based on the general description of a plasma flow with varying ion-charge composition due to step-by-step ionization \cite{shalashov_jetp_2016, abramov_rf_2016,shalashov_os_2016}. 
%
The quasi-one-dimensional plasma flow with varying transverse cross section $\sigma(x)$ is described by ion-species densities~$n_j(x)$, charge state~$Z_j=j$ (e.~g., charge~$je$), and flow velocity~$u(x)$. Assuming the electron temperature $T_e\sim100\;$eV (relevant to the experiment), one may neglect the electron-ion energy transfer due to elastic collisions and consider a strongly nonequilibrium plasma with $T_i\ll T_e$. This allows the ion pressure to be ignored in comparison with the electron pressure. At the same time, because of  high electron heat conductivity $\chi_e\propto T_e^{5/2}$, one may ignore the electron-temperature variation within the emitting volume. Then, a stationary flow is described by the following set of fluid equations: 
\begin{eqnarray*} \label{pbal}
\dfrac{d}{dx}(\sigma n_j u)=\sigma (k_{j-1} n_e n_{j-1}- k_j n_e n_j),\quad n=\sum n_j,\\
\label{mbal}
\dfrac{d}{dx}\left(\sigma M n {{u}}^2\right)=-\sigma \dfrac{d}{dx}(n_e T_e),\quad n_e=\sum Z_j n_j,
\end{eqnarray*}
representing the ionization balance and the total electron and ion momentum balance, respectively. 
Here $M$ is the ion mass, $n$ is the total density of ions, $n_e$ is the electron density following from the quasineutrality condition, and $k_j(T_e)$ is the ionization coefficient of the $j$th ion. 

Related to the EUV source, we need a continuous solution to the fluid equations in the vicinity of the ion-acoustic transition, which  unavoidably follows from the flow expanding into a vacuum. In multicomponent plasmas, the finding of such a solution faces essential mathematical difficulties since the ion-sound velocity depends on the ion-charge distribution. Then, even location of the possible singularity along the flow is formally indefinite as it is governed by the preceding ionization dynamics. Nevertheless, this uncertainty may be overcome, and a \emph{stable, stationary, continuous, and  smooth} solution may be obtained if the fluid equations are solved with some very nontrivial boundary condition as proposed in Refs.~\cite{shalashov_jetp_2016, abramov_rf_2016}.
Recently, this technique was applied and tested in a related study of an EUV light source based on a microwave discharge in a mirror magnetic trap \cite{abramov_pop_2017}.

Let the flow expand into the solid angle $\Omega$ starting from the nozzle with diameter $d$ at $x=0$. Then the cross section of the flow may be modeled as $$\sigma(x)=\pi (d/2)^2+\Omega x^2,\quad\Omega=4\pi \sin^2(\vartheta/4),$$ where $\vartheta$ is the planar expansion angle (see Fig.~\ref{Fig01}). 
With the gas preionization in mind, we assume that plasma consists of singly ionized xenon at $x=0$. 
Together with the conditions of the smooth ion-acoustic transition and the fixed gas puff rate $F=\sigma n u=\mathrm{const}$, this allows us to set the problem in a completed form. 


The electron temperature enters the model as a parameter that may be defined from energy conservation. The power that is required to support the discharge at given $T_e$ is equal to the total power losses defined as 
\begin{equation} \label{eqP}
P=\left[ \sigma n u \:(\tfrac12{M u^2}+A \langle Z_j\rangle T_e)\right]_{L}\!\!+P_{\mathrm{ion}}+P_{\mathrm{exc}}.
\end{equation}
The first term represents the convective losses corresponding to the kinetic energy flux to the collector located at $x=L$,  $A =1+0.5\ln({T_e}/{m_e u^2})\approx3$ accounts for the ambipolar plasma potential \cite{shalashov_jetp_2016}, and $\langle Z_j\rangle=n_e/n$ is the average ion charge.
The second term describes volumetric power losses for electron-impact ionization: 
\begin{equation*} \label{pion}
P_{\mathrm{ion}}=\int_0^L\!\!  p_{\mathrm{ion}}\sigma\, dx,\quad p_{\text{ion}}=\sum_{j} E_{j} k_j n_e n_j,
\end{equation*}
where $E_{j}$ is the ionization energy  of the $j$th ion, and $ p_{\mathrm{ion}}$ is the power required to ionize a unit volume of multiply charged plasma per unit time.
The third term describes power losses due to line excitation. Formally, this term may be represented in a similar way as the ionization losses:
\begin{equation}\label{pradex}
P_{\mathrm{exc}}=\int_0^L\!\!  p_{\mathrm{exc}}\sigma\, dx,\;\; p_{\text{exc}}=\sum_{j}\sum_{h,l} \Delta E_{jhl} k_{jlh}^* n_e n_{j} ,
\end{equation}
where 
indices $h$ and $l$ numerate energy levels of the excited and ground electron configurations, respectively, for all allowed transitions in the spectrum of the $j$th ion, $\Delta E_{jhl}$ is the transition energy, $k_{jlh}^*$ is the \emph{effective} excitation coefficient, and $p_{\mathrm{exc}}$ is the \emph{effective} power density.

In rarefied plasmas, once excited, an ion spontaneously relaxes to the ground state emitting a photon, which freely escapes from the discharge volume; so, with good accuracy $p_\mathrm{exc}$ may be interpreted as an actual density of volumetric losses into solid angle $4\pi$ sr, and $k_{jlh}^*$  is the usual excitation coefficient. This is exactly the case of the optically thin plasma considered in Refs.~\cite{shalashov_jetp_2016, abramov_rf_2016,shalashov_os_2016,abramov_pop_2017}. 
However, in the present paper we are considering plasma densities at least two orders of magnitude higher than before.
The xenon jet tends to be in the opposite (optically thick) limit, at least for some spectral lines. In this case, we must take into account the possibility of radiation trapping inside the emitting volume, and then losses in a particular  line may be attributed to emission from only an outer part of the discharge volume. This involves essentially new physics of radiation losses.

To calculate $k_{jhl}^*$  corresponding to the arbitrary (optically dense or thin) plasma, we generalize a technique proposed by Holstein and Biberman \cite{biberman_1982} to account for multiple terms in the ground and excited electron configurations.
The total power loss due to line excitation is expressed as
\begin{equation}\label{pradex1}
P_{\mathrm{exc}}=\int \Big(\sum_{j}\sum_{h,l}  \Delta E_{jhl} A_{jhl} \theta_{jhl} \nv_{jh}\Big)\;dV,
\end{equation}
where again $h$ and $l$ numerate terms of the excited and ground configurations, $\nv_{jh}$ is the density of excited ions (we will use overbars and underlines to help distinguishing the lower and upper energy levels), $A_{jhl}$ is the rate of spontaneous decay (the first Einstein coefficient), and $\theta_{jhl}$ is the probability of photon escape from the discharge volume. Under assumptions that 
the frequency of spontaneous emission quantum is distributed according to the line with shape $a(\omega)$
and the spatial distribution of emitting ions is uniform,
one can determine the probability $\theta_{jhl}$ as the probability that a newly born photon will pass the characteristic linear dimension $r_j$ of the volume occupied by $j$th emitting ion: 
\begin{equation*}\label{thetadef}
\theta_{jhl}=\int a(\omega) \exp\big(-r_j \kappa_{jhl}(\omega)\big)\, d\omega,
\end{equation*}
where $\kappa_{jhl}$ is the absorption coefficient, 
\begin{equation*}
\kappa_{jhl}(\omega)=\frac{1}{4}({\gv_{jh}}/{\gn_{jl}})\lambda_{jhl}^2 A_{jhl} \nn_{jl}\,a(\omega),
\end{equation*}
$\gv_{jh}$ and $\gn_{jl}$ are statistical weights of the corresponding upper level and lower level in the $l \rightarrow h$ transition, $\nn_{jl}$ is the population of the ground configuration of the $j$th ion corresponding to the $l$th energy level, and $\omega$ is the detuning from the line central frequency. The photon escape probability is strongly dependent on the mechanism of line broadening. In our case, the line form factor $a(\omega)$ must be calculated as a convolution of the Doppler and natural lines (the Voigt integral). In calculations we use approximate formulas for $\theta _{jhl}$ as a function of $r_j \kappa_{jhl}(0)$ developed by Apruzese exactly for Voigt-broadened lines \cite{apruzese_1985} .

To define the density of excited ions, we consider the stationary equation of population balance for a \emph{fixed upper energy level} under the assumption that only transitions from the ground configuration are possible (no further excitation of already excited states): 
\begin{equation}\label{bibholst}
\dfrac{d \nv_{jh}}{d t}=\sum_l \big(\kn_{jlh} n_e \nn_{jl}- (\kv_{jhl} n_e+A_{jhl}\theta_{jhl})\,\nv_{jh}\big)=0,
\end{equation}
where $\kn_{jlh}$ and $\kv_{jhl}$ are the electron-impact excitation and deexcitation coefficients for the $l \rightarrow h$ and $h\rightarrow l$ transitions, respectively, and $\nn_{jl}$ and $\nv_{jh}$ are the densities of the ground and excited configurations.

The rate of spontaneous decay is
\begin{equation*}
A_{jhl}= \frac{2e^2}{m_ec^3}\left(\frac{\Delta E_{jhl}}{\hbar}\right)^{2}\frac{\gn_{jl}}{\gv_{jh}}f_{jlh},
\end{equation*}
where $f_{jlh}$ is the oscillator strength.
The excitation coefficients, averaged over the Maxwellian electron distribution, may be estimated within the Bethe approximation for dipole-allowed transitions \cite{NRL}:
\begin{equation*}
\kn_{jlh}=  \frac{6e^4}{m_e^2c^3}\left(\frac{2\pi m_ec^2}{3T_e}\right)^{3/2}\!\frac{\exp({-\varepsilon_{jhl}})}{\varepsilon_{jhl}}\langle G(\varepsilon_{jhl})\rangle f_{jlh},
\end{equation*}
where $\langle G(\varepsilon_{jhl})\rangle$ denotes the thermally averaged Gaunt factor, and $\varepsilon_{jhl}=\Delta E_{jhl}/T_e$. 
From the principle of detailed balancing, the deexcitation coefficient is  
\begin{equation*}
\kv_{jhl}=\kn_{jlh}(\gn_{jl}/\gv_{jh})\,\exp(\varepsilon_{jhl}).\end{equation*}
Using the three equations above  together, one can calculate the number of deexcitation acts per spontaneous decay transition as
\begin{equation*}\label{betadef}
\beta_{jhl}\equiv\frac{\kv_{jhl}n_e}{A_{jhl}}=
3.7\cdot10^{-13}\;\frac{n_e  [\text{cm}^{-3}]}{T_e^{7/2}[\text{eV}]} \; \frac{\langle G(\varepsilon_{jhl})\rangle}{\varepsilon_{jhl}^3  }.
\end{equation*}
It is important to note that $\beta_{jhl}$ does not depend on any line characteristic except the transition energy. Also, for case studied, $\beta_{jhl}$  is small compared with unity for the most-important lines.  

The transition energy $\Delta E_{jhl}$ and the energy of free electrons are much higher than the difference between energy levels of the ground configuration; thus, the following condition is valid for any pair of levels $l$ and $l'$:
\begin{equation}\label{therm}
|\Delta E_{jhl}-\Delta E_{jhl'}|\ll\Delta E_{jhl},\Delta E_{jhl'},T_e.
\end{equation}
Physically, it may be interpreted as lower levels are in contact with an effective thermal reservoir. Then the levels in the ground configuration are populated proportionally to the statistical weights:
\begin{equation*}
\nn_{jl}=\alpha_{jl}\, n_j,\quad \alpha_{jl}=\gn_{jl}\,\Big(\textstyle\sum_{l'}\gn_{jl'}\Big)^{-1}.
\end{equation*}
where ${n}_j$ is the total density of unexcited $j$th ions, which is approximately equal to the total $j$th ion density. 
This allows us to solve Eq.\;\eqref{bibholst} for the population of excited levels as
\begin{equation*}\label{njh1}
\nv_{jh} = \frac{\sum_{l'} \alpha_{jl'} \kn_{jl'h}  }{\textstyle\sum_{l'} A_{jhl'}(\theta_{jhl'}+\beta_{jhl'})}\,{n_e n_{j}}.
\end{equation*}
Substituting this solution into Eq.\;\eqref{pradex1}, we find the effective volume density of power loss is
\begin{equation}\label{pradex2}
\frac{dP_{\mathrm{exc}}}{dV}= \sum_{j}\sum_{h,l}  \frac{\Delta E_{jhl} A_{jhl} \theta_{jhl}\sum_{l'} \alpha_{jl'} \kn_{jl'h}}{\sum_{l'} A_{jhl'}(\theta_{jhl'}+\beta_{jhl'})}\,{n_e {n}_{j}}.
\end{equation}
When calculating the sum over the ground levels, one must keep in mind condition \eqref{therm}. From this condition, we have $\Delta E_{jhl}\approx\Delta E_{jhl'}$ and  $\beta_{jhl}\approx\beta_{jhl'}$ (i.e., both quantities may be treated as independent of $l$). Then changing of the summation order over $l$ and $l'$ in Eq.\;\eqref{pradex2} gives 
\begin{equation}\label{pradex3}
\frac{dP_{\mathrm{exc}}}{dV}= \sum_{j}\sum_{h,l}  \alpha_{jhl}\eta_{jh} \Delta E_{jhl} \kn_{jlh}n_e n_j
\end{equation}
with
\begin{equation}\label{pradex4}
\eta_{jh}=\frac{\sum_{l'} A_{jhl'} \theta_{jhl'}}{ \sum_{l'} A_{jhl'}({\theta_{jhl'}+\beta_{jhl'}})}.
\end{equation}
After averaging over the jet cross section $\sigma$, we find these formulas result exactly in Eq.\;\eqref{pradex} with $k_{jhl}^*\equiv \alpha_{jhl}\eta_{jh}\kn_{jlh}$.

Parameter $\eta_{jh}$ has a clear physical interpretation: for each isolated transition $h\to l$ in Eq.\;\eqref{pradex2}, $\eta_{jh}$ is the ratio of actual losses to possible radiative power losses if the trapping effect is neglected. Consequently, for optically thin lines, $\beta_{jhl}\ll\theta_{jhl}$, the trapping parameter $\eta_{jh}$ is close to unity. In the opposite limiting case of optically thick lines, $\beta_{jhl}\gg\theta_{jhl}$, the trapping parameter vanishes: $\eta_{jh}\ll1$. In other words, this parameter takes into account the 
effective decrease of the region contributing to losses of ($jh$) line; such volume is defined by the condition $$\sum_l A_{jhl}\theta_{jhl}(\mathbf{r}) \gtrsim\sum_l A_{jhl}\beta_{jhl}.$$ 
Note that this condition is independent of $l$ (i.e., transitions to different ground levels $l$ corresponding to the same excited level $h$ become trapped simultaneously).

In the following, we define the fraction of power emitted in the target EUV band. It may be characterized by the EUV conversion efficiency defined as 
$$\mathrm{CE}=P_\mathrm{EUV}/P,$$ 
where $P_\mathrm{EUV}$ is calculated with Eq.\;\eqref{pradex2}, in which only transitions in the \eband band are taken into account in the summation over $(h,l)$, while the sum over $l'$ is done over all allowed transitions. Alternatively, one may use Eq.\;\eqref{pradex3} with  $\eta_{jhl}$ redefined such that only transitions in the EUV band are taken into account in the numerator of Eq.\;\eqref{pradex4}, while all other sums are calculated for all allowed transitions.

\begin{figure*}[t]
\includegraphics[width=175mm]{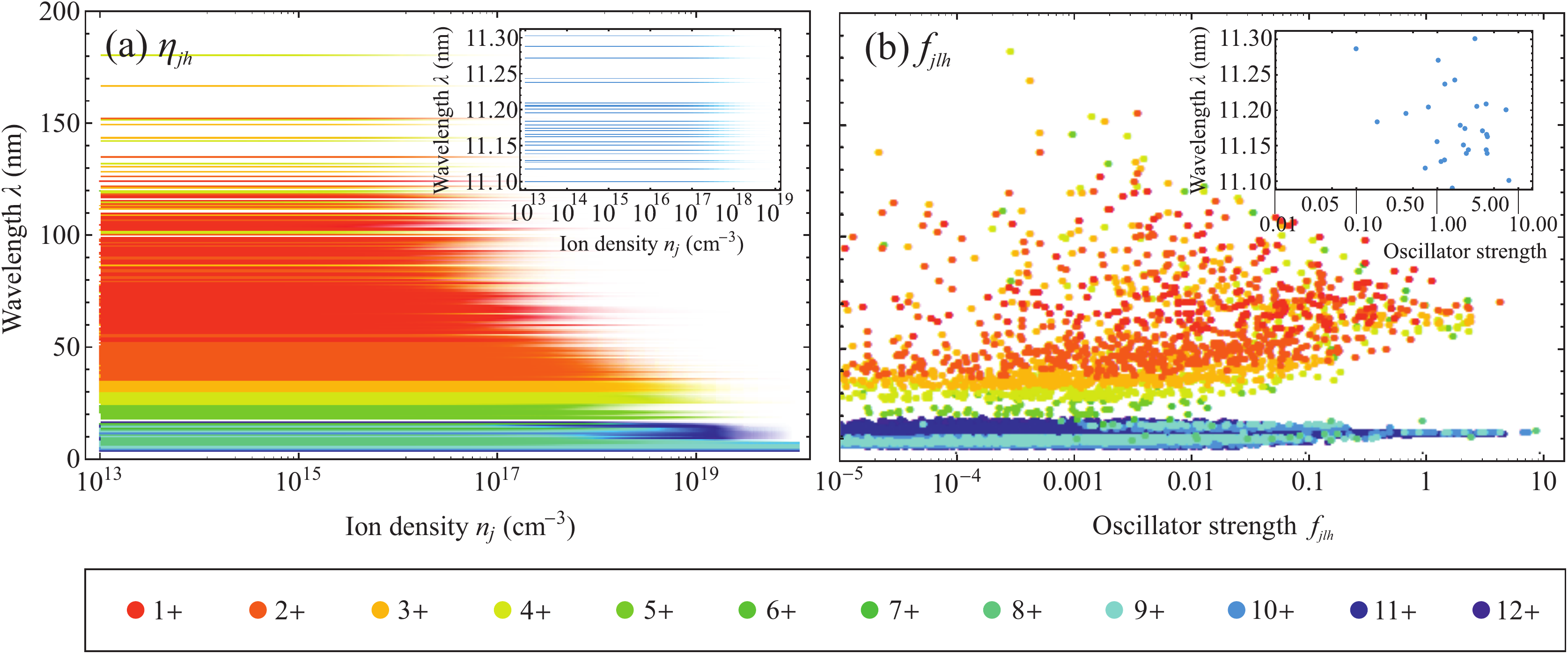}
\caption{(a) Dependence of the trapping parameter $\eta_{jh}$ on the density $n_j$ of the excited $j$th ion: the ordinate indicates the transition wavelength $\lambda_{jhl}$, different ion charges $j$ are shown with different colors, with brightness proportional to the value of $\eta_{jh}$, maximum intensities correspond to optically thin lines with $\eta_{jh}\approx1$, and vanishing colors correspond to trapped lines with $\eta_{jh}\ll1$.  (b) Oscillator strengths $f_{jlh}$ corresponding to the same transition wavelengths. The inset shows Xe$^{10+}$ lines in the \eband band in more detail. The calculation parameters are $T_e=50$~eV, $n_e=Z_j n_j$, and the characteristic linear dimension of the discharge is 100~$\mu$m. }\label{Fig02}
\end{figure*}

\section{\label{sec:atom}Atomic data for xenon}

The width of the considered EUV band of \eband\!\! is comparable to the difference between terms in the ground and excited electronic states. Therefore, the complete structure of the electron configuration for xenon ions should be taken into account.

To calculate electron-impact excitation cross sections we use the Bethe approximation; thus, for the input, we need oscillator strengths, energy levels, and statistical weights. For \eband lines, all data needed are taken from Refs.~\cite{churilov_2002,churilov_2004}, essentially based on high-resolution spectral measurements in the deep-ultraviolet region for Xe$^{9+}$ and Xe$^{10+}$. For all other lines, the spectral data are calculated with Cowan's code \cite{cowan_1981}. 
In the modeling, only allowed transitions are taken into account as they have much higher probabilities than the forbidden transitions, and the spectra contain a significant quantity of $\Delta n=0$ resonance lines that correspond to a situation when the allowed transitions dominate.

In the absence of any information concerning the electron distribution function in the discharge, the cross sections are averaged over the Maxwellian velocity distribution. We use the averaged Gaunt factor as formulated by Van Regemorter \cite{van_regemorter_1962} with the correction of Sampson and Zhang \cite{sampson_1992}.

Features of $\text{Xe}^{1+}-\text{Xe}^{12+}$ spectral lines
are illustrated in Fig.~\ref{Fig02}. Every line is characterized by its wavelength $\lambda_{jhl}$, oscillator strength $f_{jlh}$ and trapping parameter $\eta_{jh}$ for a range of emitting-ion densities. The ion spectrum shifts toward the EUV region with increase of charge. For ion densities up to $10^{17}$~cm$^{-3}$, the plasma is almost optically thin for EUV radiation. At the same time, radiation of many strong lines with smaller transition energies (longer wavelengths) may be trapped in the discharge volume, which is beneficial for the conversion efficiency into the EUV band (near 10~nm). The optimal value of the ion density in the discharge for EUV light generation seems to be somewhere in between $10^{17}$ and $10^{18}$~cm$^{-3}$, where EUV (including \eband) lines are almost the sole contributors to radiation loss. The densities above 10$^{18}$~cm$^{-3}$ are not expedient for EUV light generation, at least for the  typical example of a discharge with characteristic dimensions of about 100~$\mu$m considered. A similar result was obtained independently for laser-produced plasma by Izawa et al. \cite{izawa_2008}, who studied \uband radiation sources based on tin-droplet evaporation. Thus, our conclusion on the upper limit for density, related to EUV light trapping, seems to be general.

Monoenergetic cross sections of electron impact-ionization are taken from Ref.~\cite{povyshev_2001}, and then averaged over the Maxwellian  electron distribution function. 

\section{\label{sec:cons}Constraints for EUV sources based on a dense-xenon-plasma jet}

\begin{figure*}[tb]
\includegraphics[width=175mm]{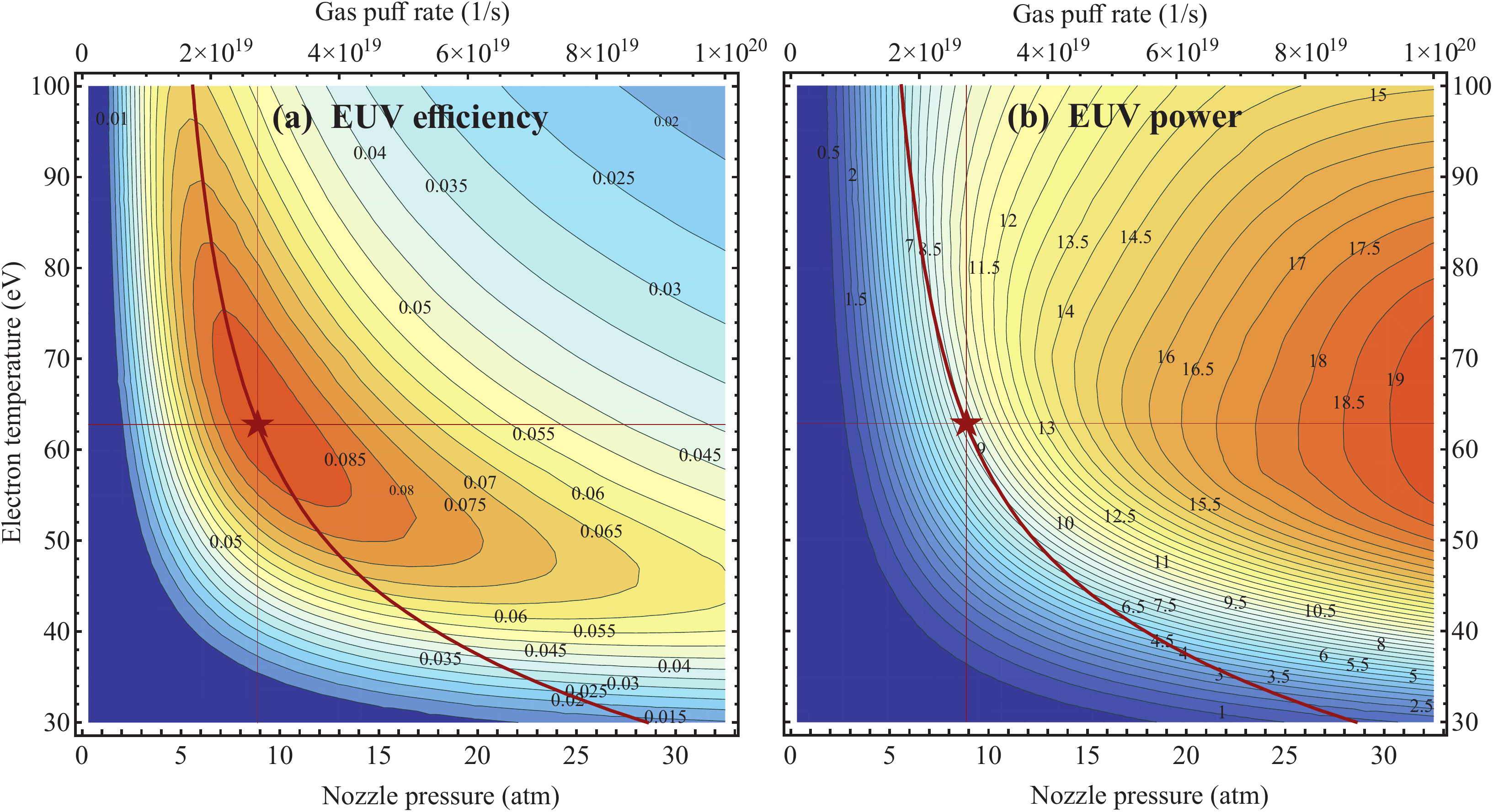}
\caption{EUV conversion efficiency (a) and EUV power (b) for the \eband band as functions of electron temperature $T_e$ and pressure at the nozzle output $p$  (or, equivalently, gas puff rate $F$) for constant nozzle diameter $d=30\,\mu$m, jet divergence $\vartheta=\pi/2$, and nozzle-to-collector distance $L=5\,$cm;   $\bigstar$ indicates the maximum conversion efficiency at $F=2.7\times10^{19}\,$s$^{-1}$ and $T_e=63\,$eV; the bold red curve crossing $\bigstar$ corresponds to constant total power load $P=100\,$kW. 
}\label{Fig03}
\end{figure*}
\begin{figure}[bt]
\includegraphics[width=80mm]{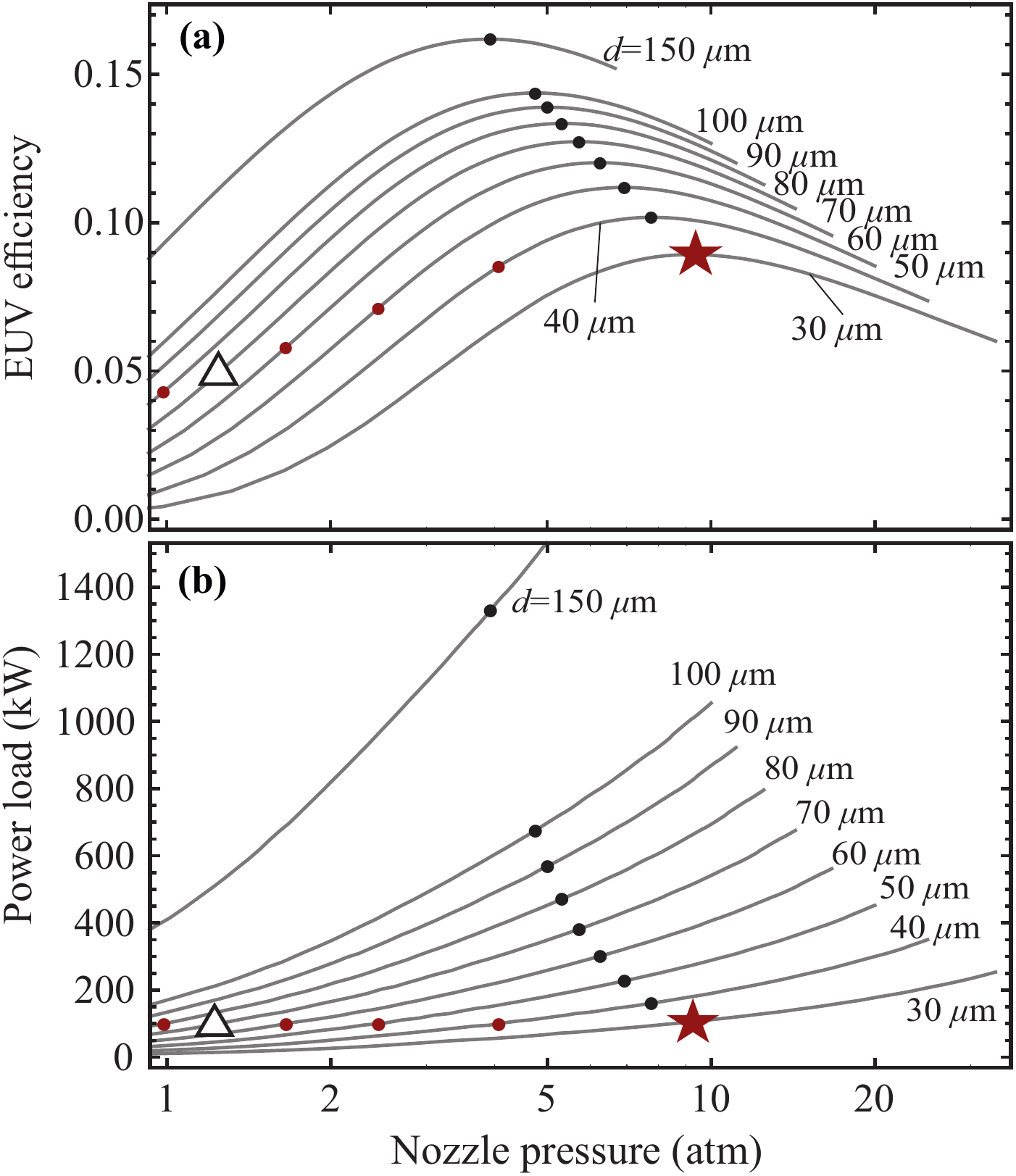}
\caption{EUV conversion efficiency (a) and required total power load $P$ (b) as  functions of pressure $p$ at the nozzle output for different nozzle diameters $d$; $\bigstar$ and black points indicate the maximum conversion efficiency and red points correspond to fixed power load $P=100\,$kW. The parameters are the same as  in Fig.~\ref{Fig03}.
}\label{Fig03b}
\end{figure}

The model described  is used to simulate the setup sketched in Fig.~\ref{Fig01}.
The control parameters studied are as follows: 
\begin{itemize}
    \item The nozzle diameter $d$.
    \item The gas puff rate which is equal to the conserved total particle flux, $F=\sigma n u$.
		\item The electron temperature $T_e$. 
\end{itemize}
In experiments, the gas puff rate is tuned by adjustment of the pressure at the nozzle output. 
Thus, to simplify a link to experiments, along  with the gas puff rate we use the reference nozzle pressure $p=T_0 F/(\sigma_0 v_0)$ calculated under the assumption that a gas with the room temperature $T_0=300\,$K flows with sound velocity $v_0=({{\frac53}T_0/M})^{1/2}$ through the nozzle cross section $\sigma_0=\pi d^2/4$ and providing the flux $F$. The electron temperature is controlled with the microwave power. Assuming that all microwave power is absorbed by electrons, the electron temperature may be treated just as a measure of the total power load into the discharge; that is,  $P=P(T_e)$ is a known function given by Eq.~\eqref{eqP} for the fixed nozzle diameter  and gas puff rate.

Before we give the numerical results, it is worth describing the underlying physics. An efficient EUV light source must satisfy two conditions: 
\begin{itemize}
    \item[(1)]  The plasma consists of a significant number of Xe$^{10+}$ ions since exactly these particular ions emit radiation in the \eband EUV band.
    \item[(2)]  The emitting volume is optically thin (i.e., the plasma density and discharge dimensions are limited to avoid the radiation-trapping effect in the EUV band).
\end{itemize}
To meet these requirements, we vary $F$, $T_e$ and $d$. 

An increase of the gas puff rate (by increase of the nozzle pressure) leads to increase of both electron average density and ion average density in the discharge. The higher the electron density, the more efficient is the electron-impact ionization and excitation, but the higher is the deexcitation rate. The increase of ion density results in more emitters and, at the same time, more absorbers of the line radiation (more efficient trapping). As a result of a neat balance between emission and absorption, there is an optimal value of the gas puff rate at which the conversion efficiency into EUV light is maximal.

The higher the gas puff rate is, the greater is the power required to support a particular electron temperature $T_e$. As the excitation and ionization rates strongly depend on the electron temperature, the microwave power defines whether the needed  Xe$^{10+}$ ions are born in the discharge or not, and determines the efficiency of their line radiation. But electron temperatures that are too high cause the significant convective power losses as the flow velocity increases (which scales approximately as the ion-acoustic velocity $v_{a}= \sqrt{T_e\langle Z\rangle/M}$). At the same time,  as the total flux $F=\sigma n u$ is conserved, the ion density decreases, resulting in decrease of the emissivity. A balance between these effects defines the optimal electron temperature or the optimal power load.

On the other hand, the nozzle diameter affects the optimal power and the optimal plasma flux, both  increasing with the diameter. Thus, one may adjust these values for practical requirements by varying the nozzle diameter. Such adjustment of the nozzle diameter  may be beneficial also for the optimization of EUV conversion efficiency. The convective power losses are proportional to $d^2$,  while the volumetric power losses, presumably due to line emission for \emph{optically thin} Xe$^{10+}$ plasma, increase as $d^3$. So, the conversion efficiency tends to increase with the nozzle diameter until the plasma is optically thin for EUV light. 

The above statements are illustrated in Fig.~\ref{Fig03}. Here the EUV conversion efficiency and EUV power are plotted as a function of the nozzle gas pressure (gas puff rate) and the electron temperature for a fixed nozzle diameter. One can see that there is a combination of the gas pressure and microwave power providing the maximum conversion efficiency (marked with ``$\bigstar$''). At this point, the total power load is $P\approx100\,$kW, the EUV power in the \eband band is 9 kW, and the electron temperature is 63 eV. 
Variation of the gas puff rate with maintenance of the same total power would result in the red hyperbola-like curve shown in Fig.~\ref{Fig03}. 

In Fig.~\ref{Fig03b}, we show the EUV conversion efficiency and corresponding total power load into electrons as a function of the gas pressure for different nozzle diameters. The conversion efficiency slowly rises with increase of the diameter  and saturates at about 15\% when the emitting region becomes large enough to switch on the essential trapping of EUV radiation. At the same time, the power required to support the optimal discharge (with the maximal conversion efficiency) increases sharply with the diameter (see the series of black points in Fig.~\ref{Fig03b}). 
Thus, the total deposited power determines the optimal nozzle diameter---it should be  as wide as possible while being consistent with the available power (see  ``$\bigstar$'' in Fig.~\ref{Fig03b}). Such fully optimized conditions may result in undesirably high gas fluxes and plasma densities in a practical design of the EUV light source. In this case, it is possible to reduce the pressure as compared with the optimal pressure and sacrifice EUV efficiency (see the series of red points corresponding to the same power load).

The power deposited into the discharge is ultimately determined by the microwave power delivered from a gyrotron. Most advanced gyrotrons are developed for magnetic nuclear fusion researches \cite{Thumm,Den}. Such devices are able to provide  megawatt-level power in long pulses (up to 1000 s) or cw mode at selected frequencies between 60 and 170 GHz. The cutoff plasma density for the electromagnetic wave at 170\;GHz is  $3.6\times10^{14}\,$cm$^{-3}$, which is  {four orders of magnitude} less than a typical electron density inside the emitting zone of the xenon jet. This is not extremely critical because of the following:
\begin{itemize}
    \item As noted before, the microwave absorption may be shifted towards lower densities in the expanding jet.
    \item For small compared to the wavelength and smoothly inhomogeneous plasma droplets, the effective absorption is possible for much larger densities than the formal cut-off density (strictly defined for a plane wave in infinite medium), see, e.g., \cite{IEEE} and references therein.
\end{itemize}
However, there are many reasons for going to higher heating frequencies, among them are the following:
\begin{itemize}
    \item Possibility to focus the microwave beam to a smaller volume.
    \item A big distance between the heating and emitting zones is prone to additional power losses and plasma instabilities.
    \item At nearly atmospheric pressure, microwave breakdown and plasma ramp-up at initial stages of the discharge go more easy at higher heating frequencies; for our conditions, this statement is confirmed experimentally \cite{sidorov_smp_2017,vodopyanov_IRMMW_2017}.
\end{itemize}
Few high-power gyrotrons operating at frequencies of 250--670 GHz are being developed \cite{subTHz1,subTHz2,Thumm}. These devices, providing peak power on the order of 100 kW level in a pulsed mode with the potential of operating in cw mode, are likely most suitable for the proposed EUV-source concept.
 

\section{\label{sec:sim}Example of an EUV source with a 200-kW, 250-GHz gyrotron }

The optimal point  in Figs.~\ref{Fig03} and \ref{Fig03b} (indicated by a star) corresponds to a feasible experiment aimed at demonstration of a physical prototype of an EUV light source with a xenon jet, which is in preparation at the IAP RAS. The experiment is arranged with a recently developed gyrotron operating at 250~GHz \cite{subTHz3}. This tube is designed for cw operation at an output power of 200~kW. Presently it works in a pulsed mode (the pulse length is up to 50$\,\mu$s with a  repetition rate 10 Hz) due to limitations of the available power supply. In  pulsed mode the gyrotron provides a peak power of up to 330~kW.

Because electrodynamic issues are outside the scope of this paper, we just summarize our present understanding. On the basis of  techniques developed in Refs.~\cite{IEEE,OS}, we estimate the maximal fraction of electromagnetic radiation absorbed by the plasma as 10--50\%. The uncertainty depends on the plasma density distribution and overall discharge dimensions (up to few mm). 

Thus, in the most optimistic case, the power load into the electron component is about 100\;kW. 
This value dictates the optimal nozzle diameter, 30$\;\mu$m. The maximum EUV conversion efficiency in the \eband band is 9\% (or 4.5\% if defined as a fraction of the total microwave power from the gyrotron) is achieved with a puff rate of $2.7\times10^{19}$~s$^{-1}$. As already mentioned, such a source provides 9~kW of EUV light when operating at an electron temperature of about 60~eV and with cold ions. The expected pressure inside the nozzle is 9~atm, which is acceptable from a technological point of view. 

\begin{figure}[tb]
\includegraphics[width=80mm]{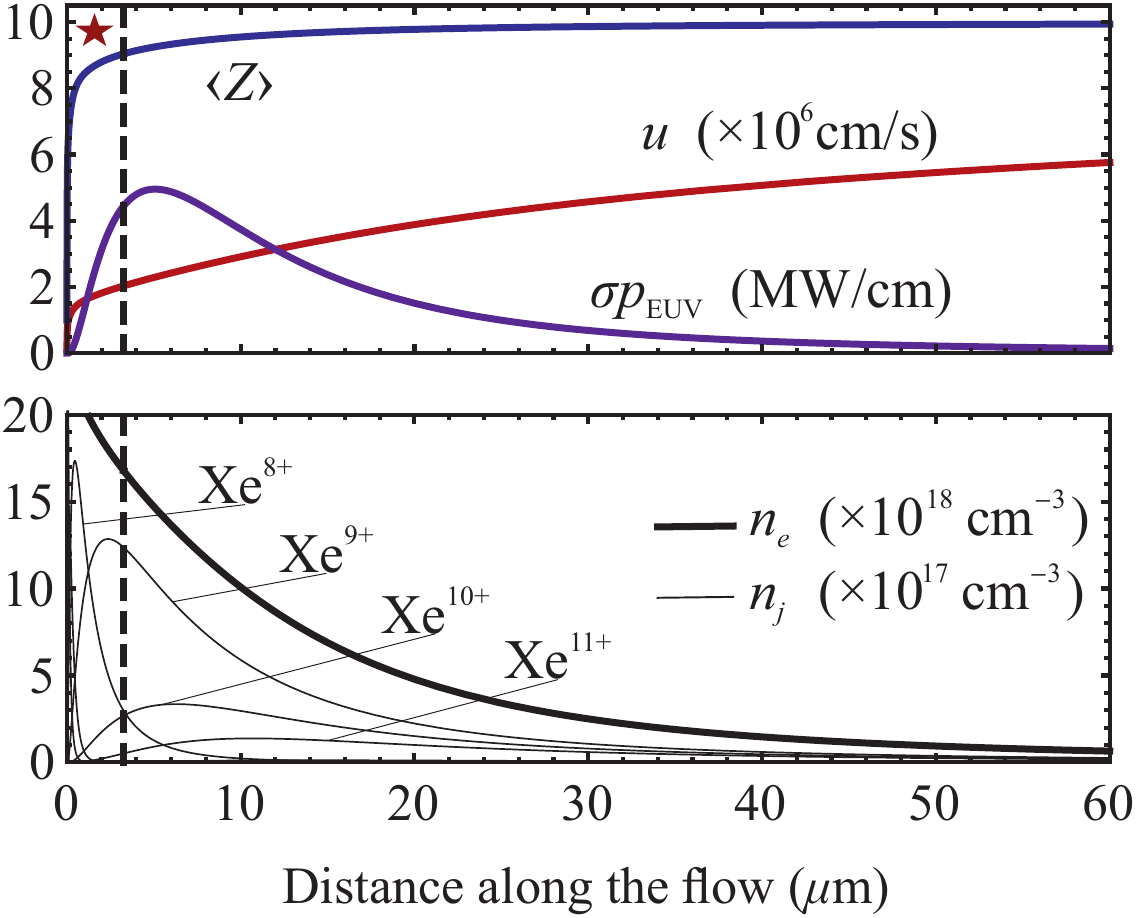}
\caption{Discharge parameters as a function of coordinate $x$ along the flow (distance from the nozzle output). Top: linear density $\sigma p_{\text{EUV}}$ of the power losses in the \eband spectral region (violet line),  plasma flow velocity $u$ (red), and average ion charge $\langle Z\rangle$ (blue). Bottom: electron density $n_e$ (thick line) and densities of ion species $n_j$ (thin lines). The parameters correspond to the point indicated by $\bigstar$ in Fig.~\ref{Fig03b}.
}\label{Fig04} 
\end{figure}
\begin{figure}[tb]
\includegraphics[width=80mm]{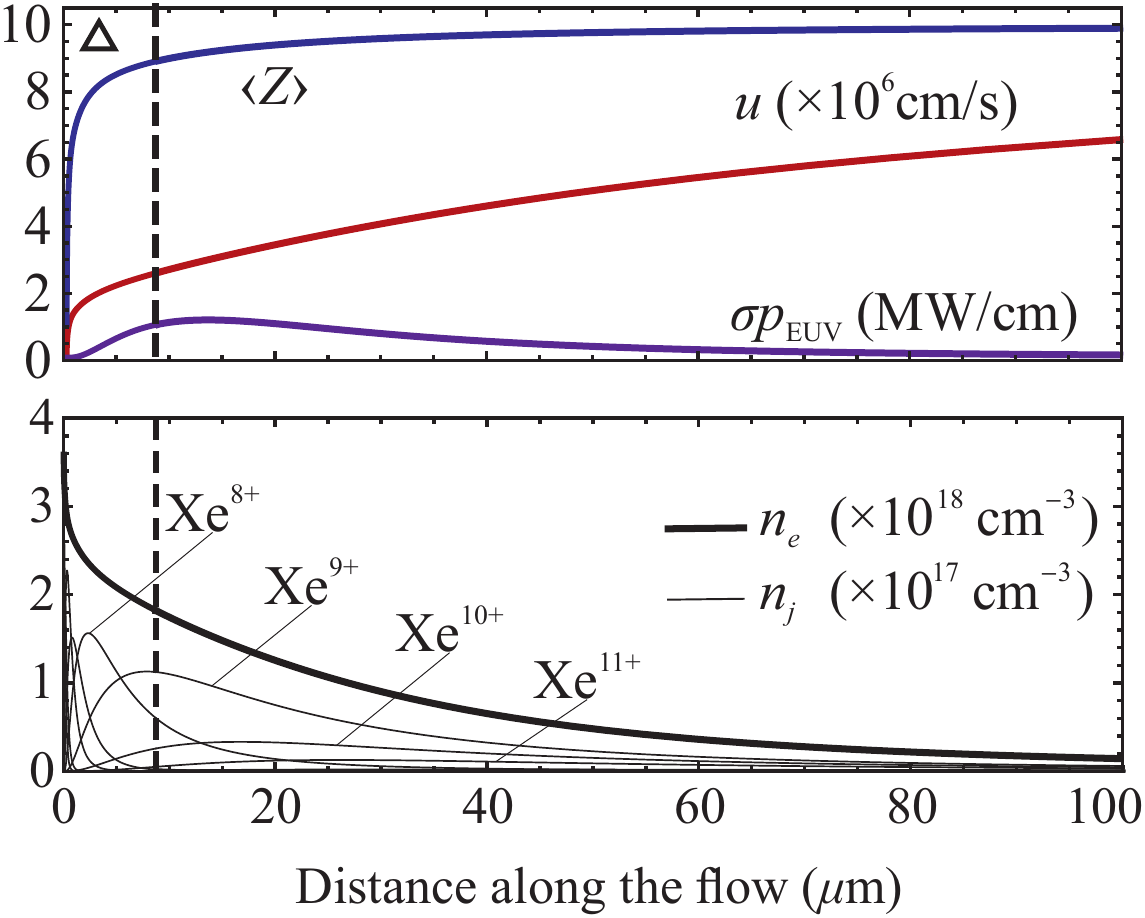}
\caption{Same as Fig.~\ref{Fig04} except the nozzle diameter $d=70\,\mu$m. The parameters correspond to the point indicated by $\triangle$ in Fig.~\ref{Fig03b}.
}\label{Fig05} 
\end{figure}

Distributions of discharge parameters along the jet are shown in Fig.~\ref{Fig04}. The maximum emissivity in the EUV band corresponds to the distance from the nozzle output $x=5\;\mu$m; this characterizes the linear dimensions of the emitting spot. This point corresponds to the ion-acoustic transition $u=v_a$, which actually is a result of the optimization since after this point the ion charge cannot increase substantially \cite{shalashov_jetp_2016}. Inside the EUV emitting zone, the electron density and average ion charge are $1.7\times10^{19}\,$cm$^{-3}$ and $\langle Z\rangle=9$, respectively. The emitting zone repeats the shape of the Xe$^{10+}$ density as this ion has the strongest lines at \eband\!\!. After the sonic barrier, the average ion charge is nearly constant, while all densities fall due to the geometric divergence and further accelerating of the plasma flow.  The cutoff plasma density is about 1~mm from the nozzle, which  may roughly characterize the largest possible dimension of the microwave absorption region.

As high initial pressure and subsequent high electron and ion densities may potentially lead to technical difficulties, at least in a proof-of-principle experiment, we also mention the not-optimal case indicated by ``$\triangle$'' in Fig.~\ref{Fig03b}. Here we assume a nozzle twice as wide with the diameter 70$\;\mu$m and keep the same power deposited into the discharge. This change essentially relaxes the conditions for the pressure and density---the  pressure inside the nozzle and the electron density reduce to 1.2~atm and $1.8\times10^{18}\,$cm$^{-3}$, respectively. The EUV conversion efficiency also reduces, to 5\%, but not drastically. Such a source provides about 5~kW of EUV light when operating at an electron temperature of about 100~eV and gas puff rate $F=2\times10^{19}\,$s$^{-1}$. The spatial distributions of the discharge parameters along the jet are shown in Fig.~\ref{Fig05}; except for different vertical scales, they are very similar to those shown in Fig.~\ref{Fig04}.

\begin{figure}[tb]
\includegraphics[width=80mm]{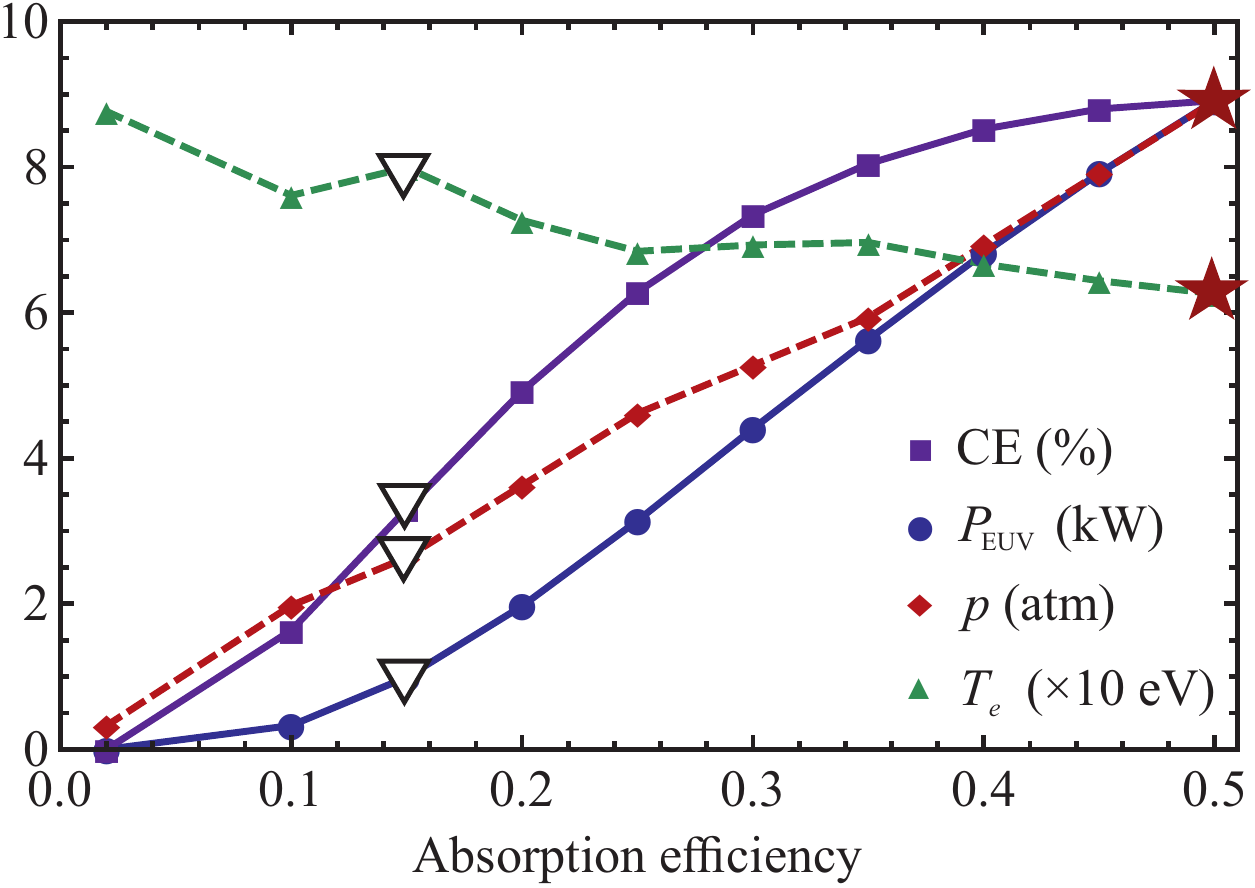}
\caption{EUV conversion efficiency (CE), EUV power $P_\mathrm{EUV}$, optimal nozzle pressure $p$, and electron temperature $T_e$ versus microwave absorption efficiency $P/P_0$ (defined as the ratio of varied loaded power $P$ to the gyrotron power $P_0=200$ kW). The gas puff rate corresponds to maximal EUV power; the other parameters are as in Fig.~\ref{Fig03}.
}\label{Fig07} 
\end{figure}

Finally, let us consider the case when the fraction of microwave radiation absorbed by the plasma falls below the optimistic value of 50\%. The corresponding degradation of EUV power and efficiency is illustrated in Fig.~\ref{Fig07}, where the abscissa shows the loaded power as a fraction of injected power provided by the 200-kW gyrotron, the nozzle diameter is 30$\;\mu$m, and the gas puff is adjusted in a such a way that the EUV power takes the maximal possible value at each point. A decrease of  loaded power leads to a almost linear decrease of the EUV power. To keep the optimal (for EUV) high ion charge with a reduced power load, one should lower the plasma density  
by decreasing  the gas pressure. One can see that, with a small enough nozzle, an EUV power of 1 kW, which is still of great interest for applications, is possible with a microwave absorption efficiency about 15\%; see the points indicated by ``$\triangledown$'' in Fig.~\ref{Fig07}. The characteristics of this regime are as follows: the nozzle pressure is 2.6~atm, the gas puff rate is $8.1\times10^{18}$~s$^{-1}$, and the electron density inside the emitting region is $4.2\times10^{18}\,$cm$^{-3}$, the electron temperature is 80~eV. 

\section{Conclusions}

We consider the concept of an EUV light source based on the line emission of multiply charged xenon ions formed and supported in a plasma jet by microwave radiation of high-power gyrotrons. Theoretical modeling, partially confirmed in preliminary experiments at a density level two orders of magnitude lower than discussed here, shows the possibility to obtain  EUV conversion efficiencies comparable to or even exceeding the efficiencies of analogous laser-produced plasma devices. Radiation trapping,  a new physics aspect considered in the present paper, may affect the energy balance in the dense xenon jet and the EUV conversion efficiency significantly, but this influence is mostly positive as radiation of unwanted lines with low transition energies is trapped in the discharge volume at lower densities than the EUV lines. The physics of formation and supporting of an EUV-radiating plasma jet appears to be clear and robust.  
In this context, we rely on ongoing experimental verification of theoretical scaling and further fast progress in the development of high-power subterahertz gyrotrons. 

\vspace{3pt}\emph{Note added after the paper was accepted.} Preliminary results of the experiment discussed in Sec.\ref{sec:sim} are now available; see Ref.~\cite{APL}.

\begin{acknowledgments}
The work was supported by the Russian Foundation for Basic Research~(Grants No. 17-02-00173 and No. 18-32-00419).
I.S.A. acknowledges support from the Foundation for the Advancement of Theoretical Physics and
Mathematics ``BASIS'' (Grant No. 18-1-5-12-1). 
The authors thank S. V. Golubev and A. V. Sidorov for inspiring discussions and comprehensive assistance.
\end{acknowledgments}

\end{document}